# Dislocation-point defect interaction on plasticity across the length scale in $SrTiO_3$

Chukwudalu Okafor[1*], Kohei Takahara[2], Svetlana Korneychuk[1,3,4], Isabel Huck[5], Sebastian Bruns[6], Ruoqi Li[2], Yan Li[2], Karsten Durst[6], Atsutomo Nakamura[2*], Xufei Fang[1*]

[1]Institute for Applied Materials, Karlsruhe Institute of Technology, 76131, Karlsruhe, Germany

[2]Department of Mechanical Science and Bioengineering, The University of Osaka, 560-8531 Osaka, Japan

[3]Institute of Nanotechnology, Karlsruhe Institute of Technology, 76344, Eggenstein-Leopoldshafen, Germany

[4]Institute of Nanotechnology and Karlsruhe Nano Micro Facility (KNMFi), Karlsruhe Institute of Technology, 76344 Eggenstein-Leopoldshafen, Germany

[5]Department of Chemistry, Technical University of Darmstadt, 64287 Darmstadt, Germany

[6]Department of Materials and Earth Sciences, Technical University of Darmstadt, 64287 Darmstadt, Germany

*Corresponding authors: chukwudalu.okafor2@kit.edu (C.O.); a.nakamura.es@osaka-u.ac.jp (A.N.); xufei.fang@kit.edu (X.F.)

**Abstract**

Point defect engineering is widely used to tailor the electronic and transport properties of complex oxides, yet its influence on dislocation plasticity remains poorly understood. Here, we establish how donor (Nb) doping modifies dislocation nucleation, multiplication, and mobility in single-crystal SrTiO3 by bridging nano-, meso-, and macroscale deformation. Using a combinatorial approach involving nanoindentation, cyclic Brinell indentation, and bulk uniaxial compression, we show that 0.5 wt% Nb doping consistently suppresses room-temperature plasticity. Nanoindentation reveals increased pop-in stresses, increased lattice friction stress, and reduced creep rates, indicating inhibited dislocation nucleation and motion with Nb doping. Mesoscale Brinell indentation exhibits discrete, widely spaced slip traces reflecting more difficult dislocation multiplication. Bulk uniaxial compression confirms ~50% higher yield stress in Nb-doped (0.5 wt%) $SrTiO_3$ samples. Comparison with Fe-doped $SrTiO_3$ (equivalent doping concentration) isolates the role of defect chemistry: oxygen vacancies promote incipient plasticity, whereas Sr vacancies dominate in Nb-doped $SrTiO_3$, strongly hindering dislocation motion. This length-scale bridging approach consistently reveals suppressed dislocation nucleation, multiplication, and motion in the 0.5 wt% Nb-doped samples. These insights underline the importance of dislocation-defect chemistry on the mechanical behavior of functional oxides.

## 1. Introduction

Doping is a common strategy for tuning the physical properties of a wide range of functional oxides. In particular, donor doping has been widely employed to manipulate the mobile charge carrier species and concentrations, hence impacting the electrical conductivity of oxides [1-3]. For instance, by donor doping, in $SrTiO_3$ (with perovskite oxide structure, $ABO_3$), the cations (A or B site) are substituted with a dopant with higher valency (e.g., $La^{3+}/Sr^{2+}$ on the A-site, or $Nb^{5+}/Ti^{4+}$ on the B-site), and the resulting defect chemistry depends on the temperature and oxygen partial pressure ($pO_2$) [4-6]. As such, the dominating diffusing ions (cation or anion, electrons/holes) scale with temperature, concentration of defect species, and $pO_2$ in $SrTiO_3$ and related perovskite oxides. Hence, tailoring of various physical (electrical [1] and thermal conductivity [7]) and microstructure (grain size and boundary structure [8]) is possible via donor doping. Besides point defects engineering, recently dislocations (one-dimensional line defects) have been shown to influence promising aspects for tuning both functional [9, 10] and mechanical properties of oxides [11, 12], akin to doping. For instance, Nb-doped $SrTiO_3$ has been recently demonstrated with higher superconductivity in the presence of mechanically induced dislocations [13].

Although using both dislocations and point defects to enhance the physical properties of oxide ceramics appears appealing, potential bottlenecks may be encountered when considering both defect types for their synergy in improving functionality. First, a ternary oxide such as $SrTiO_3$ has cation and anion species with varying activation energies, and the enthalpy of formation of the respective vacancy species [14, 15]. By donor doping, as in the case of Nb doping, $Nb^{5+}$ atoms substitute $Ti^{4+}$ on the B-site of the lattice, leading to electronic and/or strontium vacancy compensation depending on the ambient oxygen activity [6, 16, 17]. Second, there are other reports of possible lattice distortion [8, 18, 19], lattice expansion [19], formation of stacking faults [18] which breaks the local symmetry of the lattice, and segregation of Nb dopants to defects such as grain boundaries [20], leading to more complexities. Since engineering dislocations into these materials is a prerequisite for achieving dislocation-tuned properties, this leads to the following questions: How will these complex mechanisms impact the dislocation-based mechanical properties such as plasticity and/or cracking of Nb-doped $SrTiO_3$? If possible, would there be competing mechanisms at different length scales, analogous to the observation by tuning the non-stoichiometry of nominally undoped single-crystal $SrTiO_3$ [21]?

Bulk uniaxial compression has been, for decades, the most common mechanical testing technique employed for probing the mechanical properties of ceramics, with rich literature on $SrTiO_3$ [22-26] and other oxides [27-30]. However, the progressive miniaturization of functional devices and the advancement of nano-/micromechanical testing methods [31] demand bridging the deformation scale and unifying the understanding of the mechanical behavior of materials at different length scales, if achievable. Also, the large sample requirement for bulk uniaxial compression, particularly for single crystals, translates to a very high cost. Yet it remains a challenge to bridge the dislocation-based mechanical behavior in crystalline solids across the length and time scales. To this end, various attempts have been made. For instance, Fang et al. [32] developed an experimental protocol for probing the mechanical response of nominally undoped $SrTiO_3$ at the nano-/ and macro-scales. Similarly, Fincher et al. [33] have shown the significant strain rate sensitivity of Li metals through bulk tensile testing and nanoindentation. However, neither study accounts for mesoscale deformation, missing the detailed experimental information and mechanisms between the nanoscale and macroscale testing.

Here, we develop a multiscale deformation approach, coupled with donor (Nb) doping, to address dislocation behavior in the model perovskite oxide $SrTiO_3$ across the length scales. To be specific, we report the effect of donor doping on the dislocation mechanics (nucleation, multiplication, and motion) in single-crystal $SrTiO_3$ at room temperature. Through a combination of nano-/microscale indentation, mesoscale cyclic Brinell indentation, and macroscale bulk uniaxial compression tests, we experimentally investigated the impact of Nb-doping on the plastic deformation of this material. In short, at all length scales we consistently observed less pronounced dislocation plasticity with increased Nb concentration, visualized via coupled dislocation etch-pit method and scanning electron microscope (SEM) examination at the nanoscale and optical microscopy slip trace examination at meso-/macroscales. We also discussed the dominating dislocation mechanisms at different length scales impacted by Nb doping by considering the space charge model of dislocations in oxides.

## 2. Experimental procedure

### 2.1. Materials selection and preparation

Verneuil-grown single-crystal $SrTiO_3$ with the (001) surface orientation and a dimension of 5 mm × 5 mm × 1 mm (Alineason Technology GmbH, Frankfurt am Main, Germany) were used for the nanoindentation and Brinell indentation tests. Nominally undoped (readily acceptor-doped due to trace acceptor impurities during crystal growth [34] as commercially available) and Nb-doped (0.05 wt.% and 0.5 wt.%) crystals were used to study the effect of donor (Nb) doping on the dislocation plasticity at room temperature.

Before nanoindentation, the sample was chemically etched in 15 mL of 50% $HNO_3$ with 16 drops of 40% HF solution for ~15 seconds to reveal the pre-existing dislocation density (inset **Fig. 3**). It is a critical step to ensure that the pre-existing density is low and has the same order of magnitude (~$10^{10}/m^2$) in all used crystals, so that we can rule out the possible impact of different densities of pre-existing dislocations [35] on the observed mechanical behavior during the mechanical tests at a later stage (see **Fig. S1** for more details). This allows for direct assessment of the effect of doping (vacancies) on incipient plasticity. For nanoindentation tests, the surface roughness was evaluated using an atomic force microscope (AFM) (Vecco, Plainview, NY, USA), which gives a surface roughness of less than 1 nm.

For uniaxial bulk compression tests, bulk $SrTiO_3$ single crystals grown by the same Verneuil method, and provided by Shinkosha Co., Ltd., Japan, were used. The Nb-doped (0.05 wt%) and (0.5 wt%) samples have dimensions of 3 mm x 3 mm x 7.5 mm, with six sides carefully polished. The long axis is along the [001], and the side surfaces are parallel to the (100) and (010) planes. Before compression testing, the samples underwent sequential cleaning with acetone, ethanol, and ultrapure water to remove possible surface contaminants.

### 2.2. Mechanical testing

Mechanical testing was performed at three different length scales to observe the dislocation behavior with donor doping. As schematically illustrated in **Fig. 1**, nanoindentation was employed to study dislocation nucleation and motion at the nano-/microscale, while cyclic Brinell indentation, as a fast and reproducible experimental approach recently developed by the current authors [36], provides direct evidence of the doping effect on dislocation-based plasticity at mesoscale. In the end, we compared the experimental outcome with the macroscale bulk uniaxial compression. Bear in mind that bulk compression requires expensive single crystals in large dimensions, and by including this we attempt to close the experimental loop for bridging the length scale. Later based on the findings, we will discuss the feasibility of excluding bulk compression tests in the future whenever practical.

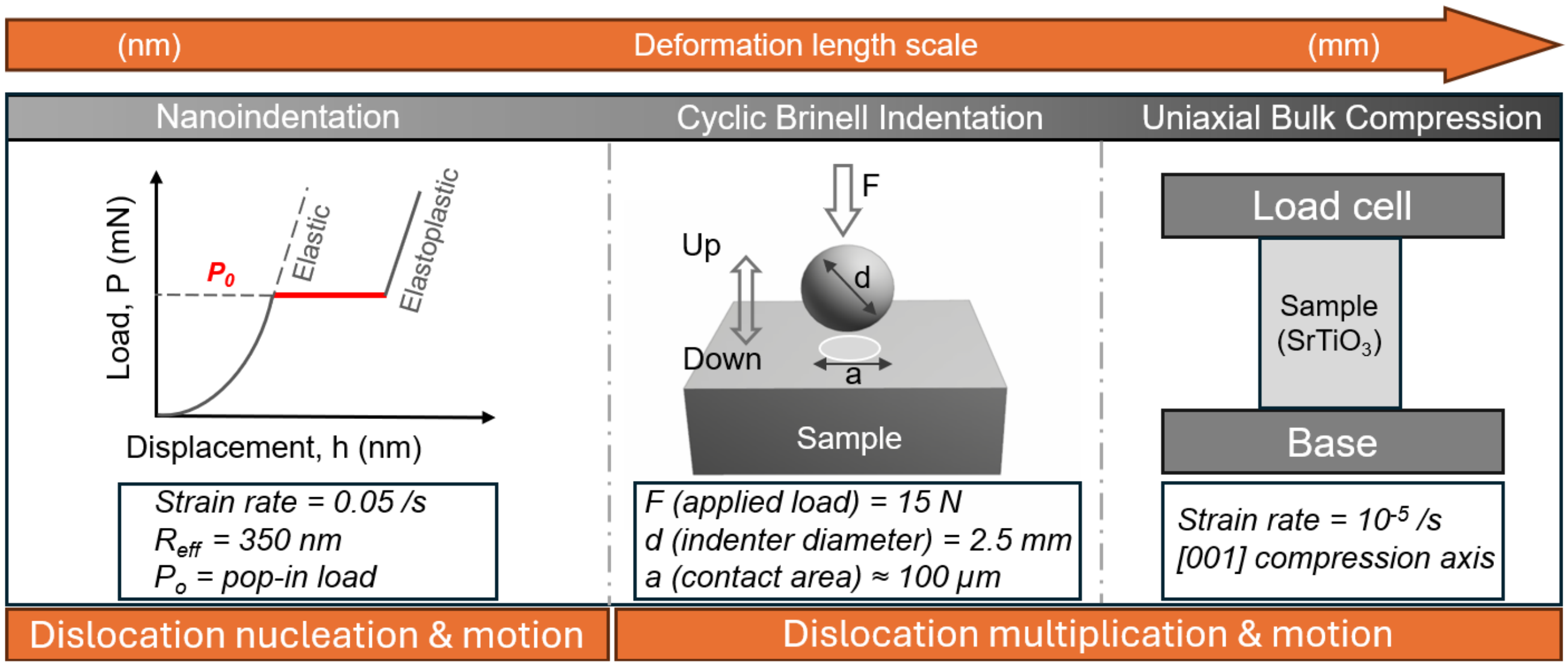

***Figure 1:*** *Experimental design across the length scale. (left) Illustration of nanoindentation (nano-/microscale) load-displacement (P-h) curves depicting the pop-in event (transition from elastic to elastoplastic deformation),*

*which allows for probing dislocation nucleation and motion. (middle) Cyclic Brinell indentation (mesoscale) and (right) bulk uniaxial compression (macroscale), addressing dislocation multiplication and motion.*

### 2.2.1. Nano-/microscale: Nanoindentation

Nanoindentation using a Berkovich indenter tip (effective tip radius of 350 nm) at a constant strain rate of 0.05 $s^{-1}$ was performed on each sample to observe the doping effect on dislocation nucleation (pop-in event) and dislocation motion (creep strain rate) [37]. A maximum displacement of 200 nm was maintained for all nanoindentation tests (except where specified), with harmonic displacement and frequency targets of 2 nm and 45 Hz, respectively, in the continuous stiffness measurement (CSM) mode using the G200 nanoindenter (KLA Instruments, California, USA). For statistical analysis, 100 indents were performed on each sample for the pop-in studies. A peak-hold time of 10 seconds was set for the nanoindentation pop-in test and 10 minutes for the creep strain rate tests.

To estimate the lattice friction stress using the dislocation etch pit approach developed by Gaillard et al. [38], we choose a maximum penetration depth of 700 nm to introduce a large plastic zone with more than 20 dislocations per etch pit arm [39]. For reproducibility, 25 indents were performed for each test condition.

### 2.2.2. Mesoscale: Brinell cyclic indentation

We employed a universal hardness testing machine (Karl-Frank GmbH, Weinheim-Birkenau, Germany) equipped with a Brinell indenter to investigate the impact of doping on dislocation multiplication at the mesoscale. Cyclic Brinell indentation was performed on all samples using a hardened steel spherical tip with a diameter of 2.5 mm and a load of 1.5 kgf [36]. The samples were secured to a precision-fitted detachable holder (with an error margin of ±1 μm), which allows the capture of the impression and slip trace evolution after each indentation cycle using an optical microscope (ZEISS Axio Imager 2, Carl Zeiss Microscopy GmbH). Note that 1x and 10x represent 1 and 10 indentation cycles, respectively. The evolution of the slip traces after each cycle, up to 10 cycles, was better enhanced in the circular-differential interference contrast (C-DIC) mode on the microscope. The procedure is repeated on at least 10 different positions for statistical analysis. A laser scanning microscope (Keyence VK-9700, Keyence Corporation, Osaka, Japan) was used to quantify the change in depth profile after indentation (**Fig. S4**).

### 2.2.3. Macroscale: Uniaxial bulk compression

At the macroscale, a universal testing machine (AGX-V 50 kN, Shimadzu, Japan) was used to perform uniaxial bulk compression experiments on the Nb-doped (0.05 wt%) and (0.5 wt%) samples. The samples were compressed along the [001] axis at a constant strain rate of $1 \times 10^{-5}\, s^{-1}$ at room temperature.

### 2.3. Dislocation structure characterization

A second round of chemical etching is performed to reveal dislocations etch pits within the plastic zone after nanoindentation. The dislocation etch pits were captured in scanning electron microscopy (SEM) (Merlin Gemini 2, Carl Zeiss Microscopy GmbH) at an acceleration voltage of 5 kV. Transmission electron microscope (TEM) characterization was carried out with ThermoFischer Scientific Themis Z equipped with SuperX EDX detector at 300 kV on the reference and Nb-doped (0.5 wt.%) samples with 10x indentation as two extreme cases for comparison. EDS analysis was performed for elemental composition. Annular dark-field (ADF) STEM images were acquired with a 145 mm camera length and about 40-200 mrad collection angles.

## 3. Results and Analyses

We begin with cyclic Brinell indentation, which offers a fast and cost-effective approach with reproducibility probing dislocation plasticity at mesoscale. Subsequently, nanoindentation and uniaxial bulk compression results were presented at the nano-/microscale and macroscale, respectively, bridging the deformation length scale with a focus on the effect of donor doping on dislocation mechanics.

### 3.1. Mesoscale cyclic Brinell indentation

By adopting the cyclic Brinell indentation method [36], we probe the mechanical response of Nb-doped $SrTiO_3$ at the mesoscale. **Figures 2a-c** feature the representative optical images after a single indentation cycle (1x) on the reference and Nb-doped samples. A discrete distribution of the slip traces with Nb-doping is evident, which is more pronounced on the Nb-doped (0.5 wt.%) sample. Further indentation in the same position up to 10x is displayed in **Figs. 2d-f**. Irrespective of the doping concentration, the slip trace density increases compared to 1x. However, the slip trace density increases after 10x is more evident for the reference sample than for the two Nb-doped samples. It is striking that only a few more slip traces were observed after 10x on the Nb-doped (0.5 wt.%) sample, as highlighted with arrows in **Fig. 2c & f**. Although the slip trace direction seems unsymmetrical for the Nb-doped (0.5 wt%), repeated cyclic Brinell indentation (**Fig. S2**) on different positions on the sample rules out possible unsymmetrical slip trace distribution. However, the discrete slip trace distribution remains consistent. With a confocal laser microscope, we quantified the profile of the impression after 10x Brinell indentation cycles with profile depths of ~360 nm, ~290 nm, and ~250 nm for the reference, Nb-doped (0.05 wt%), and Nb-doped (0.5 wt%) samples, respectively (for more details see **Fig. S3**).

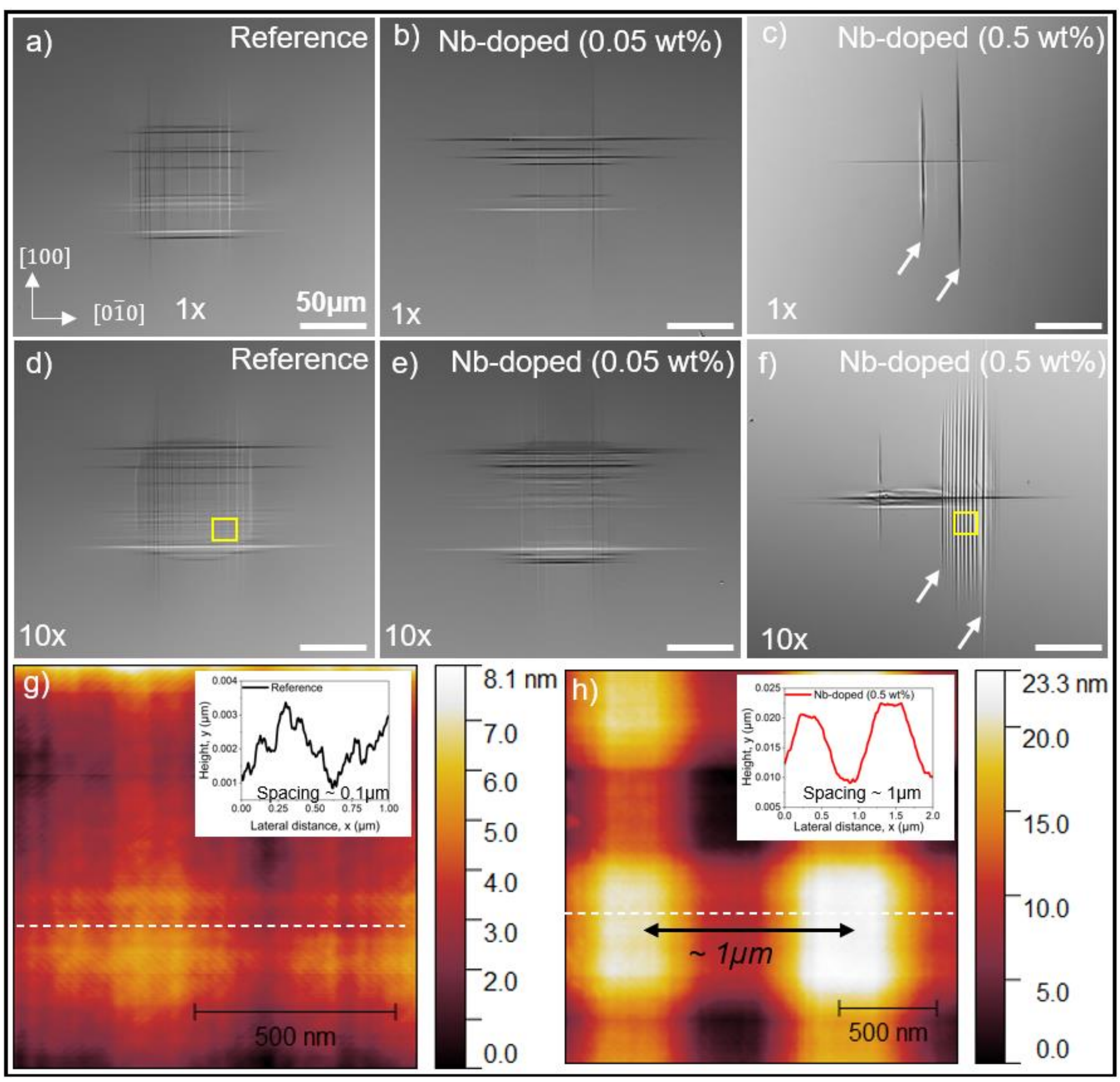

***Figure 2****: Optical microscope images depicting the slip traces after Brinell indentation (a-c) after a single indentation (1x), (d-f) after 10x indentation cycles on the same positions as in (a-c). Atomic force microscope (AFM) images were captured on the regions highlighted in yellow and presented in g and h for the reference and Nb-doped (0.5 wt%) samples, respectively. Insets in g and h show topographic information depicting the spacing between individual slip steps.*

Aside from the discrete nature of the slip traces on the Nb-doped samples, the spacing between individual slip traces is more pronounced than in the reference sample. Atomic force microscope (AFM) characterization (**Fig. 2g & h**) estimated a spacing of ~0.1 μm and ~1 μm for the reference and Nb-doped (0.5 wt%) samples, respectively. This one-order-of-magnitude difference hints at possible difficulties in dislocation multiplication and, in turn, sluggish dislocation motion due to Nb-doping.

### 3.2. Nanoindentation pop-in and creep tests

#### *3.2.1. Dislocation nucleation*

To understand dislocation nucleation at the nanoscale, we adopt the nanoindentation pop-in method [40, 41]. A representative indentation load-displacement *(P-h)* curve of the reference, Nb-doped (0.05 wt.%), and Nb-doped (0.5 wt.%) is presented in **Fig. 3a- c**. The elastic part of the *P-h* curve overlaps irrespective of doping concentration, as fitted using the Hertzian elastic contact theory [42] $P = \frac{4}{3} E_r\sqrt{R}h^{\frac{3}{2}}$. Here $P$ is the load at the pop-in event (burst in displacement at constant load on the *P-h* curve), $R$ the effective tip radius of 350 nm fitted using the Hertzian elastic contact theory [42], $E_r$ is the reduced modulus, which is defined by $\frac{1}{E_r} = \frac{(1-v_s^2)}{E_s} + \frac{(1-v_i^2)}{E_i}$. Subscripts $i$ and $s$ represents indenter and sample, respectively. With $E_i$ = 1140 GPa, $v_i$ = 0.07, $E_s$ = 264 GPa and $v_s$ = 0.237 for $SrTiO_3$ [24] an $E_r$ value of 225 GPa was calculated.

The distinct difference in the pop-in load with doping is reflected on the cumulative probability plot of the maximum shear stress ($\tau_{max}$) presented in **Fig 3d**. $\tau_{max}$ estimated using the relation $\tau_{max} = 0.31 \left(\frac{6E_r^2}{\pi^3 R^2} P_{pop-in}\right)^{\frac{1}{3}}$. A higher $\tau_{max}$ was observed for the Nb-doped samples compared to the reference sample. An average $\tau_{max}$ value of ~10.5 GPa was calculated for the Nb-doped (0.5 wt.%) sample, which is slightly higher than the Nb-doped (0.05 wt.%) and the reference samples with average $\tau_{max}$ values of ~10.4 GPa and ~9.8 GPa respectively.

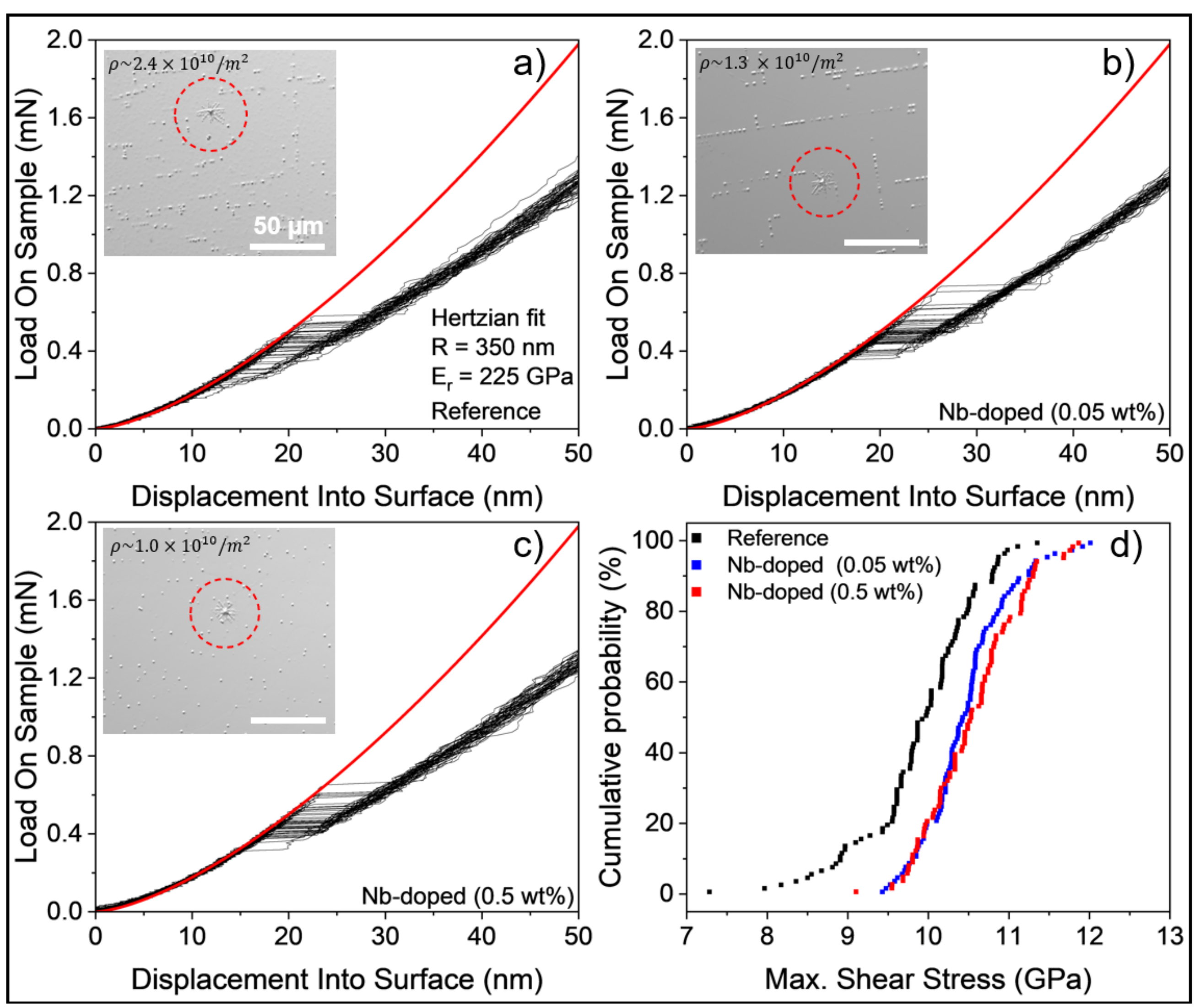

***Figure 3**: P-h curves for (a) reference [43], (b) Nb-doped (0.05 wt.%) [43], and (c) Nb-doped (0.5 wt.%). (d) The cumulative plot of the maximum shear stress for dislocation nucleation. Inset in a-c is the optical microscope image after chemical etching. Single indents were intentionally captured and highlighted in red dashed circles. Note: ρ is the dislocation density.*

### *3.2.2 Dislocation motion*

While the pop-in event during nanoindentation is attributed to dislocation nucleation, the motion of the dislocations is also important, as the newly generated dislocations must be sufficiently mobile at room temperature to avoid pile-up and subsequent cracking. Here, we analyzed the dislocation motion following two different approaches. First, lattice friction stress is quantified by the dislocation pile-up model in nanoindentation tests [38]. Second, the nanoindentation creep test estimates the creep strain rate as a function of time during the load-hold phase of the nanoindentation test (performed on the same samples as in **Fig. 3**).

To validate and quantify the impact of Nb doping on the dislocation motion, we examined the dislocation etch-pit spacing after nanoindentation testing by adopting the dislocation pile-up model proposed by Gaillard et al. [38] as presented in **Fig. 4a - f**. The model was proposed based on the principle that residual impression results in dislocation spacing during unloading. This spacing is a representative of the balance between the lattice friction stress ($\tau_f$), the interaction between dislocations ($\tau_d$), and the residual stress field of the indenter ($\tau_a$), which is given by the relation; $\tau_f = \tau_a + \tau_d + \tau_{im}$ at the equilibrium condition. Here $\tau_{im}$ is the surface image stresses acting on the dislocations. For edge dislocations, $\tau_{im} = 0$. $\tau_d$ is thus estimated in Equation (1) as:

$$\tau_d = \frac{G\vec{b}}{2\pi}\sum_{i\neq j}^{N}\frac{A}{r_i - r_j} \tag{1}$$

where $N$ is the number of dislocations on a single pile-up arm, $A$ is 1 for screw dislocations and $^1/_{1-\nu}$ for edge dislocations, with $\nu$ being the Poisson's ratio. At the plateau, the lattice friction stress $\tau_f$ is approximately equal to the $\tau_d$, as previously validated on nominally undoped single-crystal $SrTiO_3$ by Javaid et al. [39].

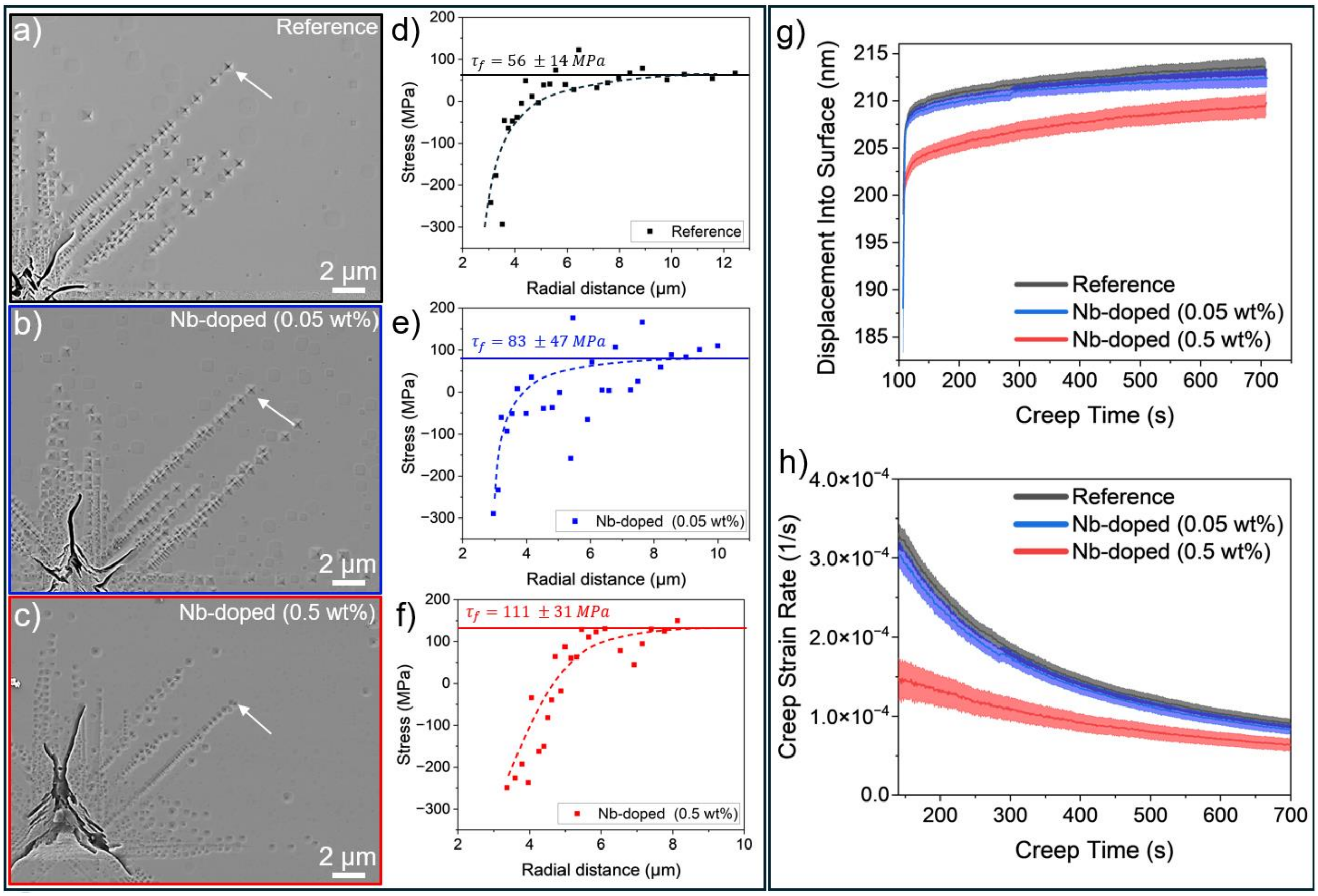

***Figure 4**: Impact of Nb-doping on dislocation motion in single crystal $SrTiO_3$. SEM images after chemical etching on (a) reference, (b) Nb-doped (0.05 wt%), and (c) Nb-doped (0.5 wt%). White arrows highlight dislocation etch-pit pileup on the <110> arm, corresponding to the arm used to estimate the lattice friction stress. Corresponding shear stress curves in (d-f) show the lattice friction stress for the respective samples. Dashed curves on (d-f) are trend lines serving as a guide. (g-h) Nanoindentation creep strain as a function of time and displacement.*

*(g) Displacement into the surface at load-hold as a function of time. (h) Nanoindentation creep strain rate estimated from the displacement in (g), as a function of time. The shaded area is the standard error from averaging 12 different tests for each sample condition.*

Note that only isolated dislocation pile-up arms are considered to avoid the influence of parallel dislocation pileups. The lattice friction stress estimated by simple averaging of the outermost dislocations, where the shear stresses are almost constant (at the plateau, using the dashed lines as guide, **Figs. 4d-f**), with values of 56 ± 14 MPa, 83 ± 47 MPa, and 111 ± 31 MPa for the reference, Nb-doped (0.05 wt%), and Nb-doped (0.5 wt.%) samples, respectively, as presented in **Figs. 4d-f**.

In **Figs. 4 g & h**, plots of displacement and indentation creep strain rate as a function of time. The shaded area on each plot is the standard error from averaging 12 tests for each sample condition. With a holding time of 10 minutes, starting at the maximum displacement of 200 nm, a distinct difference in material displacement with doping concentration is observed in **Fig. 4g**. A slight increase in displacement, which saturates with time, is visible for all samples. The indentation creep strain rate [44] in **Fig. 4h** is defined as $\varepsilon = \frac{\Delta h}{h} = \frac{h-h_0}{h}$, where $h_0$ is the initial displacement (200 nm following the experimental condition), and $h$ is the instantaneous displacement. The strain rate is hence estimated as $\dot{\varepsilon} = \frac{\varepsilon}{t}$. Initially, the strain increased abruptly with time, followed by a quasi-steady creep behavior where the strain increased linearly. A higher creep strain rate is observed for the reference sample, which is slightly lower for the Nb-doped (0.05 wt.%) sample and strongly reduced for the Nb-doped (0.5 wt.%) sample. The indentation creep behavior suggests that dislocations are more sluggish in the 0.5 wt.% Nb-doped sample.

### 3.3. Macroscale bulk compression

The uniaxial bulk compression test is a common mechanical testing approach for ceramics, with extensive literature on oxides including $SrTiO_3$ [22, 25]. This method is also seen as an ultimate validation tool for room-temperature bulk plasticity in ceramics. To this end, we present (**Fig. 5a**) the true stress-strain curve for reference, Nb-doped (0.05 wt.%) and Nb-doped (0.5 wt.%) after uniaxial bulk compression. Analogous to the nano-/microscale, there is an overlap of the elastic part for all samples just before yielding, hinting at the insignificant influence of doping on the elastic modulus. However, the onset of yielding varied significantly with doping concentration. Almost identical strength at yield for the reference and Nb-doped (0.05 wt.%) (~112 MPa and ~102 MPa, respectively), while a ~50% higher yield strength (~170 MPa) was observed for the Nb-doped (0.5 wt.%) sample. As expected, all samples show a continuous increase in the stress after yielding and attaining a fracture strain of 13.6%, 12.9%, and 13.1% for the reference, Nb-doped 0.05 wt%, and 0.5 wt%, respectively. The plastic yield in bulk compression is associated with dislocation motion and multiplication in materials. Regardless of the length scale and change in stress states between bulk compression and the indentation method, all the tests unanimously suggest that the dislocations in 0.5%wt Nb-doped samples are more difficult to move and multiply, as will be discussed in detail in **Sec. 4.2.2**.

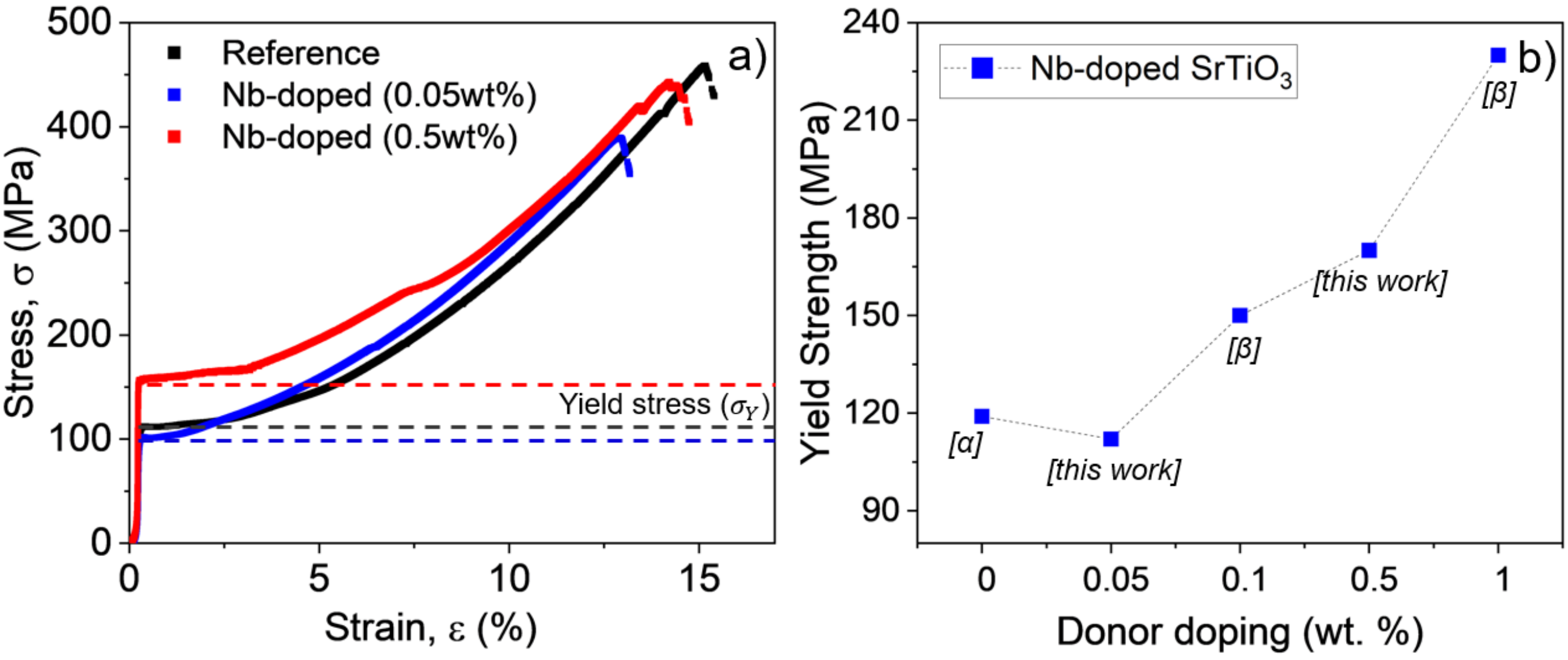

***Figure 5****: (a) Stress-strain curves of reference [25], Nb-doped (0.05 wt.%), and Nb-doped (0.5 wt.%) single-crystal samples deformed at a constant strain rate of 1.0 x $10^{-5}$ $s^{-1}$ during uniaxial bulk compression. Dashed lines indicate the yield stress, and the fracture strain is highlighted with an arrow on the blue plot (Nb-doped (0.05 wt%)). (b) Yield strength plotted against doping concentration for reference and Nb-doped strontium titanate. Data point [a] is reproduced from Ref. [25], while the tests [this work] were performed on the same uniaxial bulk compression setup as in [a]. Data points [b] for Nb-doped 0.1 wt% and 1.0 wt% were collected from Ref. [13].*

**Figure 5b** illustrates an increase in the yield strength (from bulk uniaxial compression test) with Nb doping concentration (up to 1 wt%) for the current study and available literature data. Similar to our observation at different length scales with increasing Nb concentration, the literature data aligns with our current observations. Whether the observed decrease in plasticity with Nb-doping will continue until plasticity is completely lost (little to no plastic deformation before fracture) is beyond the scope of the current study due to the unavailability of samples with higher Nb concentration. However, it is worth noting that the solubility limit of Nb dopants in $SrTiO_3$ could be as high as ~14 wt%, which depends on the concentration of acceptor impurities and co-dopants [45].

### 3.4. Dislocation characterization

Finally, we rule out possible segregation of dopants and cations in the vicinity of dislocations and potential impact on the observed mechanical behavior of Nb-doped $SrTiO_3$, by characterizing the elemental composition of the dislocation as depicted in **Fig. 6**. Annular dark-field (ADF) image of a dislocation line on Nb-doped sample (0.5 wt%, with the most drastic change in dislocation-based plasticity (**Figs. 2f)**) is shown in **Fig. 6a**. The EDX mappings in **Fig. 6b-d** for the corresponding area in **Fig. 6a** suggest no detectable segregation of the Nb dopant or cations around the dislocations. It may be argued that this is due to the low doping concentration in the present case. However, Nb segregation was not observed with even up to 10.2 at% (≈ 5.0 wt%) Nb-doping in $SrTiO_3$, as previously reported by Rodenbücher et al. [47]. This appears not surprising as the driving force for B-site atom diffusion in $SrTiO_3$ is energetically unfavorable [46].

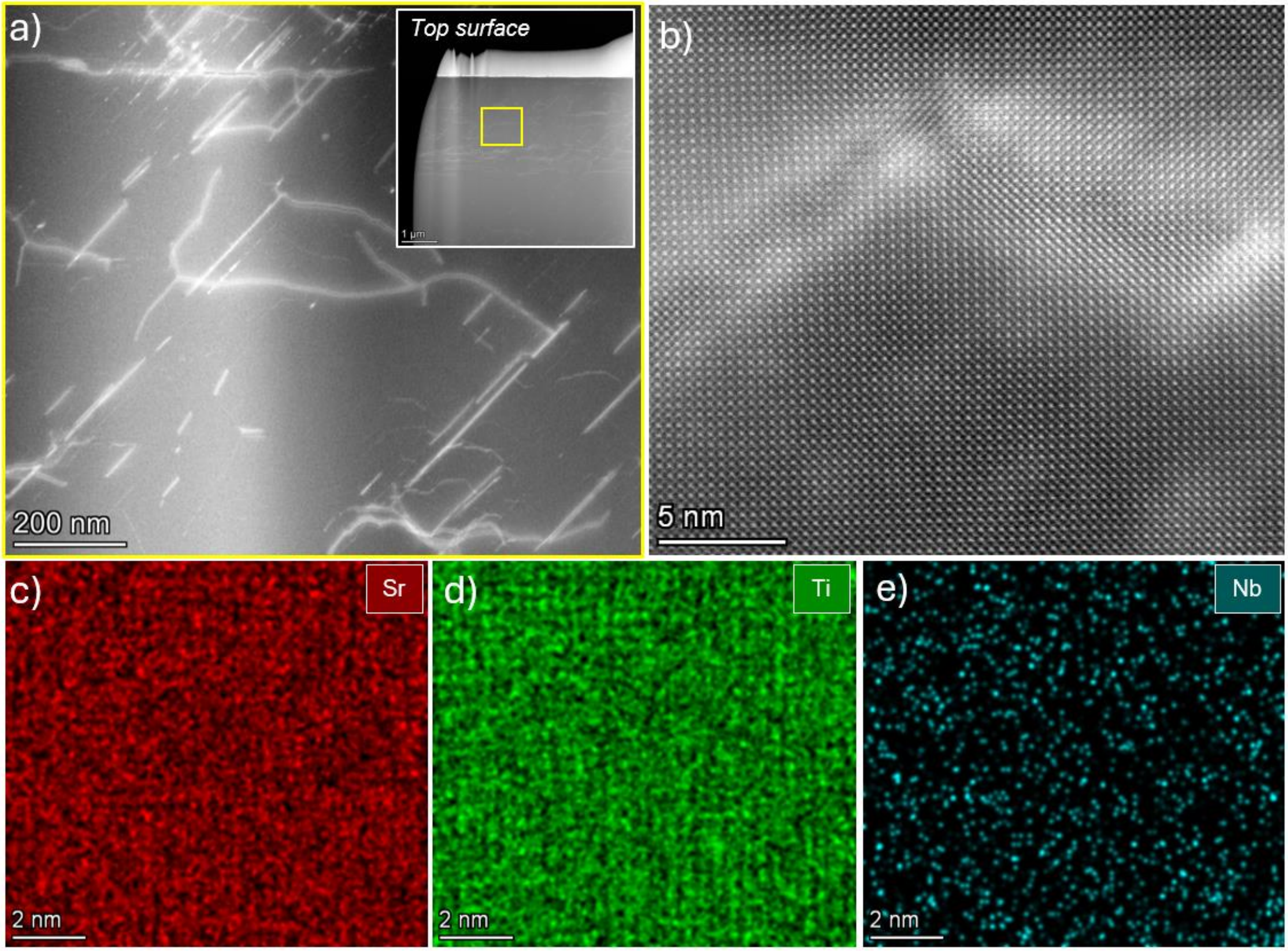

***Figure 6***: *(a) Annular dark-field imaging (ADF) of dislocation lines in Nb-doped (0.5 wt.%) after cyclic Brinell indentation. The inset shows the overview of the cross-section after FIB milling. (b)High-resolution, high-angle annular dark-field (HAADF) scanning transmission electron microscope (STEM) image of the dislocation region. Corresponding STEM EDX precisely to the field of view in Fig. (b), showing homogenous distribution and no segregation around the dislocations within the EDX detection limit of about 0.1 at%, in (c) Sr, (d) Ti, and (e) Nb.*

### 4. Discussion

When compared with dislocations in metals, the charge feature of dislocations in ceramics with ionic/covalent bonding shall be considered [47]. To understand the impact of donor (Nb) doping on the plasticity of single-crystal $SrTiO_3$, it is essential to identify and address the interactions between point defects and dislocations. Our approach here is to discuss separately (and, where necessary, synergistically) the impact of Nb doping on dislocation nucleation, multiplication, and motion [35, 48], focusing on the changes in defect chemistry caused by doping. We begin with a basic understanding of defect chemistry and the potential modifications resulting from Nb-doping, including the dominant vacancy species, the dislocation core structure, and the corresponding space charge. Subsequently, we consider the possible impact of vacancies (types and concentrations) within the stressed volume [32] and their effect at different length scales. Additionally, we further validate our findings on donor (Nb-doped) samples using acceptor-doped (Fe-doped) $SrTiO_3$ samples of equivalent doping concentration.

### *4.1. Influence of Nb-doping on defect chemistry*

We start with the consensus on the dislocation core structure and the compensating space-charge layer. Irrespective of the dislocation type (edge [49] or screw [50]), doping (acceptor or donor [51]), and method (experiment [51-53] or simulation [54]), the dislocation core in $SrTiO_3$ is found to be oxygen vacancy-rich. This is to a large extent attributed to the lower energy of oxygen vacancy formation near the dislocation core [55]. Therefore, nucleated dislocations not only sweep up vacancies during their motion, but can also acquire charges of the (oxygen) vacancies [56]. On the other hand, the diffusion of cation (Sr) vacancies at room temperature (**Fig. S3**) is negligible and thus does not form a depleting space charge around the dislocation core as schematically illustrated in **Fig. 7a** (nominally undoped) and **b** (Nb-doped).

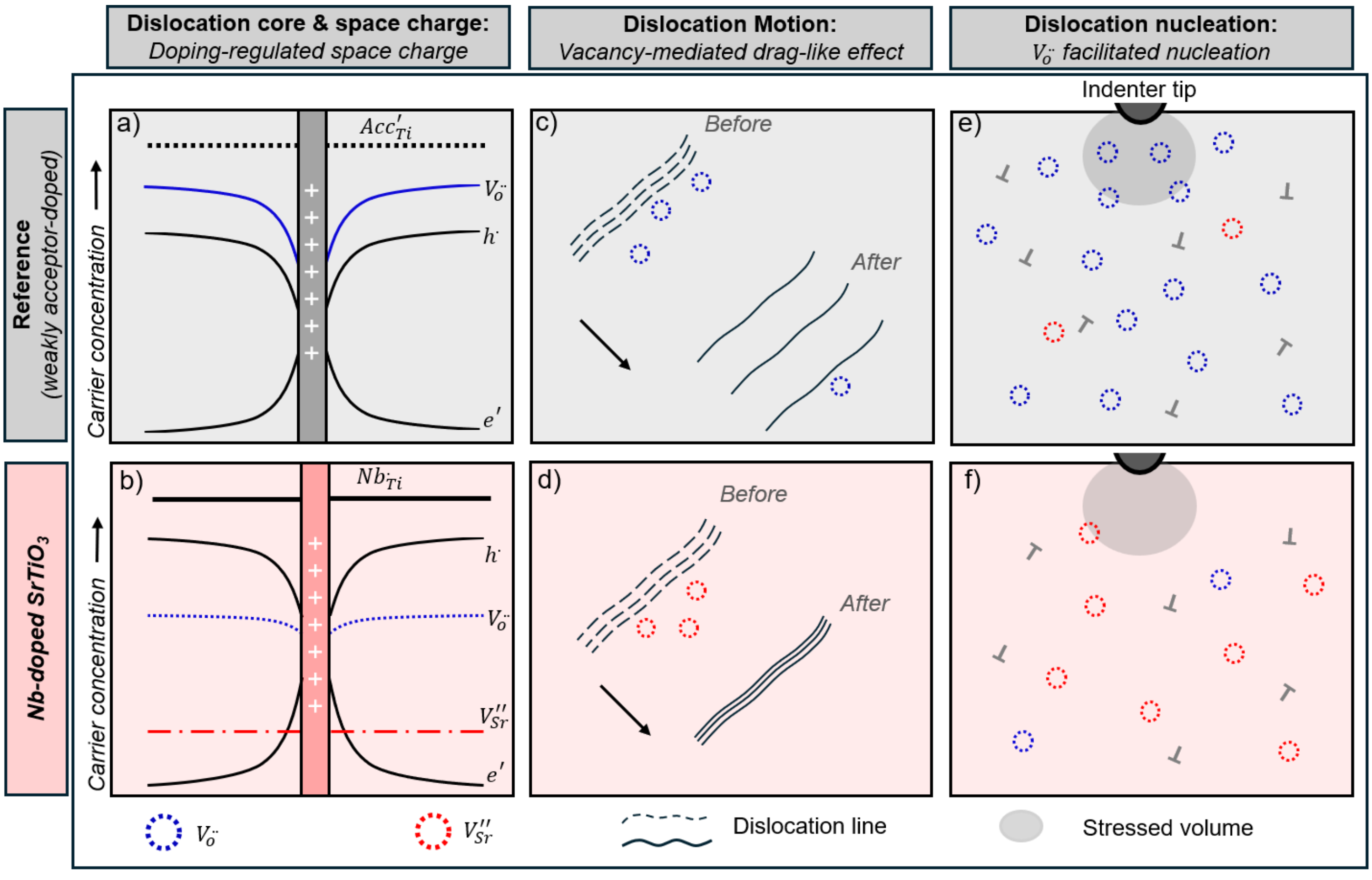

***Figure 7**: Overarching illustration of dislocation behavior influenced by donor (Nb) doped $SrTiO_3$. Left: Dislocation core structure and compensating space charge regions: (a) reference (weakly acceptor doped), (b) donor (Nb) doped $SrTiO_3$. At room temperature, the high concentration of Sr vacancies in Nb-doped $SrTiO_3$ is immobile, hence they do not segregate nor*

*form a space charge around the dislocation core. (c, d): Drag-like effect of vacancies on dislocation motion. Strontium vacancies pose more significant resistance to dislocation motion (and concurrently, dislocation multiplication). Dashed lines represent the initial position of dislocation just before the interaction of vacancies in motion, while arrows indicate the direction of dislocation motion. Dislocation nucleation facilitated by oxygen vacancies [43] in (e) reference $SrTiO_3$, and (f) Nb-doped $SrTiO_3$ where strontium vacancies dominate.*

As a direct consequence, Nb-doping alters the defect chemistry of $SrTiO_3$ by sequentially reducing the oxygen vacancy concentration with increasing Nb content and at high $pO_2$ (up to 1 bar) [6]. Note that flame fusion (Verneuil process) of single-crystal growth is performed with an oxygen source and an additional annealing in an oxygen-rich atmosphere [57], it is expected that strontium vacancies will compensate for the donor (Nb) defects [6]. Only at $pO_2 > 10^{-15}$ Pa [16] are oxygen vacancies dominating the defect profile of Nb-doped $SrTiO_3$. The single-crystal $SrTiO_3$ samples used for this study were synthesized via the Verneuil method, which involves high $pO_2$ conditions ensuring the Sr vacancy-compensated Nb-doped $SrTiO_3$. Hence, only oxygen vacancies, electrons, and electron holes contribute to the space charge region around the newly generated dislocation cores of $SrTiO_3$ **(Fig. 7a, b)** at room temperature during/after deformation. There are reports of dopant segregation to the grain boundary (and grain boundary dislocations) in perovskites [20, 58]. Still, these are reported to occur at sintering temperatures where not only dopants, but also Sr vacancies are sufficiently mobile [59]. Therefore, we will mainly consider the impact of vacancies for the interpretation of our results for the room temperature mechanically inserted dislocations and their behavior.

### 4.2. Impact of Nb-doping on the dislocation mechanics

#### *4.2.1. Dislocation nucleation*

A lower pop-in load in nanoindentation can be linked to an easier nucleation of dislocations, as it has been observed qualitatively in the vicinity of oxygen vacancies in Fe-doped and nominally undoped $SrTiO_3$ [21, 25, 26]. Recently, further experimental evidence was provided for dislocation nucleation facilitated by oxygen vacancies in nominally undoped $SrTiO_3$ during room-temperature nanoindentation coupled with electromigration tests, corroborated by molecular dynamics simulation [43]. However, doped-$SrTiO_3$ will modify the dominating vacancy type as well as its concentration, depending also on the doping concentration. Contrary to the observations on nominally undoped $SrTiO_3$, the results in **Fig. 3** show a higher $\tau_{max}$ values for dislocation nucleation in Nb-doped $SrTiO_3$. The higher $\tau_{max}$ for dislocation nucleation is a direct consequence of Nb-doping that depends on several factors: dopant concentration, dominating vacancy type, and possibly lattice strain/distortion.

Several reports of lattice strain/distortion scaling with doping (acceptor and/or donor doping) concentration in $SrTiO_3$ exist [8, 18, 60]. Lattice strain/distortion should serve as *soft spots*, facilitating dislocation nucleation, as observed in nanoindentation pop-in studies of BBC medium entropy alloys [61]. The degree of lattice expansion/distortion depends on the difference in the ionic radii [62] between the host atom and the dopant. For instance, donor (Nb) on $SrTiO_3$ leads to up to 0.1% lattice expansion for 0.5 wt% Nb [19]. However, in **Fig. 8b**, the stress required for nucleating dislocations is lower for the Fe-doped and higher for the Nb-doped sample when compared to the reference sample. If lattice distortion were to influence dislocation nucleation, the pop-in stress would exhibit a similar trend (both Fe- and Nb-doped as both increase the lattice expansion) compared to the reference sample. The different trends, however, suggests that lattice distortion should have a less significant effect compared to the vacancy types, caused by doping in $SrTiO_3$, on dislocation nucleation.

In Fe-doped $SrTiO_3$ over a wide temperature and $pO_2$ range [63, 64], oxygen vacancies dominate the defect profile. Hence, the trend in **Fig. 8b** corroborates the observation in previous studies that oxygen vacancies primarily facilitate dislocation nucleation. Consider that Nb-doping decreases significantly the oxygen vacancy concentration (up to 3 orders of magnitude, scaling with temperature, $pO_2$, and doping concentration [16]) compared to the nominally undoped and Fe-doped $SrTiO_3$, the pop-in stress increase in Nb-doped $SrTiO_3$ is most likely attributed to the lower oxygen vacancy concentration.

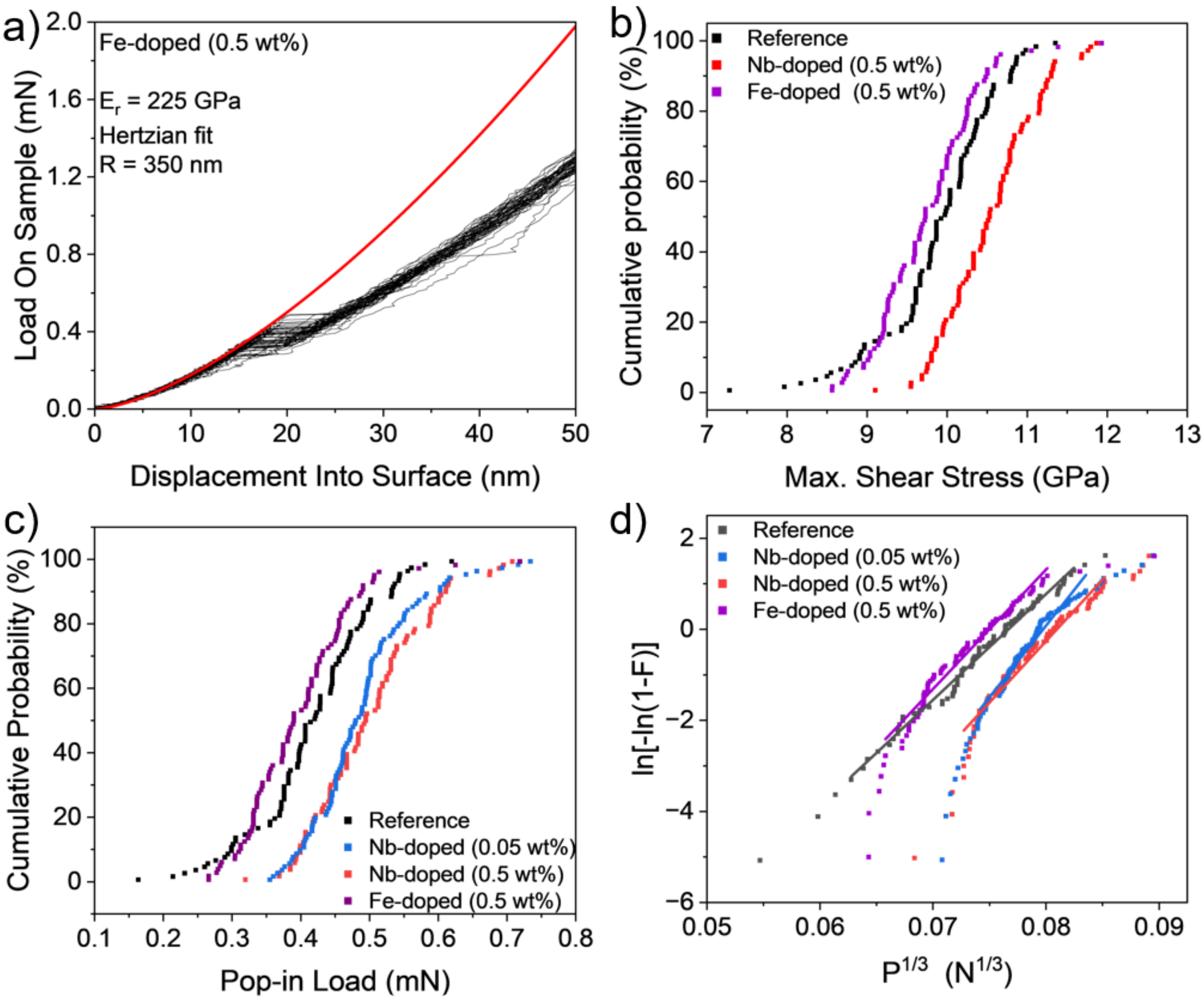

***Figure 8****: Load-displacement (P-h) curves for (a) Fe-doped (0.5 wt.%). (b) The cumulative plot of the maximum shear stress for dislocation nucleation of the reference [43], Fe-doped (0.5 wt.%), and Nb-doped (0.5 wt.%)*

*samples. (c) Recalculated cumulative probability plot for the pop-in load for the different doped samples. (d) Replot of (c) to obtain the slope (α) as in Equation 1 to estimate the activation volume.*

To shed more light on the impact of doping and doping concentration, we estimated the activation volume involved in the dislocation nucleation process [65]. Activation volume *V*, is given as $V = \frac{\pi}{0.47}\left(\frac{3R}{4E_r}\right)^{\frac{2}{3}} kT\alpha$ where, $k = 1.38 \times \frac{10^{-23} J}{K}$, $T = 303\ K$, and $\alpha$ is the slope in **Fig. 8d** (for more details, see [65]). The estimated activation volumes for reference, Fe-doped (0.5 wt.%), Nb-doped (0.05 wt.%), and Nb-doped (0.5 wt.%), are 7.20 $Å^3$, 8.12 $Å^3$, 9.74 $Å^3$, and 8.36 $Å^3$, respectively. For all samples, the activation volume is significantly less than $10b^3$, where *b* is the Burgers vector. This indicates that the deformation is dominated by homogeneous dislocation nucleation or vacancy-related heterogeneous dislocation nucleation. A slightly lower activation volume was observed in the reference sample compared to the doped samples, which suggests a possible impact of Nb-doping. For such a small activation volume, diffusion-related processes involving vacancies would dominate the deformation, as predicted by DFT modelling on $CeO_2$ [66]. This slightly higher activation volume recorded for Nb-doped samples than for the reference and Fe-doped samples suggests that dislocation nucleation in the vicinity of scarcely existing oxygen vacancies is more difficult in Nb-doped samples. We note that the activation volume could be over- (or under-) estimated following Mason's method [65] as the $V$ is dependent on the slope $\alpha$, but the general trend remains the same.

### *4.2.2. Dislocation motion*

As observed at the nano-/microscale, oxygen vacancies promote dislocation nucleation. There is, nevertheless, a reversal trend in the dislocation motion of reference (nominally undoped) $SrTiO_3$ as previously reported [21], suggesting a dual role in dislocation mechanics by oxygen vacancies. After nucleation, the oxygen vacancies are proposed to have a drag-like effect on the moving dislocations, restraining their motion analogous to the solute drag effect [67]. Higher concentration of oxygen vacancies is shown to reduce the dislocation creep rates, namely higher resistance to dislocation motion, in $SrTiO_3$ [26]. Here in the Nb-doped sample (dominating point defect species are strontium vacancies [16]), we observed a higher resistance to dislocation motion at the nanoscale as presented in **Figs. 4**. Since Nb-doped samples have lower oxygen vacancies, this suggests a different defect species is operative with a more significant impact than oxygen vacancies on hindering the dislocation motion in Nb-doped $SrTiO_3$.

To understand the impact of doping on dislocation behavior, one of the key hypotheses lies in the dislocation core. Oxygen vacancies are not only attracted to the dislocation core, but they also relax the strain energy at the dislocation core, as also reported in $CeO_2$ [68]. Upon straining during mechanical deformation, moving dislocations will interact with vacancies and affect the continuous motion of the dislocations.

The self-diffusion coefficient and, in turn, the diffusion lengths of oxygen and strontium vacancies are presented in **Fig. S3**. The diffusion length of oxygen and strontium vacancies at room temperature (after 1 second) is ~6 nm and ~$10^{-16}$ nm, following Fick's second law. In principle, the interaction of a dislocation during glide with vacancies would lead to a drag-like effect, which should scale with the mobility of the vacancy, supporting our results regarding consistently reduced dislocation motion, experimentally captured at the nano-/microscale (**Figs. 3 & 4**) and macroscale in **Fig. 5**. Additionally, there are reports of surface segregation of A-site cation vacancies in $SrTiO_3$ under an oxidizing atmosphere [16, 17]. Under an oxidizing atmosphere (> $10^{-5}$ Pa), Nb-doped $SrTiO_3$ is charge-compensated by strontium vacancies, with a gradient in the strontium vacancy concentration which inversely scales with the depth [16]. This further supports that sluggish strontium vacancies would, in principle, serve as '*stumbling blocks*' for the motion of dislocations.

Nevertheless, the stress levels required to move dislocations in $SrTiO_3$ are much lower compared to those for the dislocation nucleation [35]. With a large Brinell indenter tip of radius 2.5 mm, a contact diameter of ~100 µm, and a pre-existing dislocation density of $10^{10}$ $m^{-2}$, we can rule out homogeneous dislocation nucleation, as dislocations should be readily available within the plastic zone; hence, dislocation multiplication and motion dominate at this length scale. We adopted the indentation stress formulation by Swain et. al. [69] to estimate the mean stress during spherical indentation. Swain et al. [69] proposed an indentation stress (mean stress) estimate for spherical indentation as $p_o = {}^{P}/_{\pi a^2}$, $P$ is the applied load (15 N in the current case), and $a$ is the contact radius (~50 µm). A mean stress of ~1.9 GPa is estimated for a single indentation cycle. Note that the maximum resolved shear stress on the activated slip planes is $0.47p_o$, equivalent to ~900 MPa.  This stress level is one order of magnitude lower than the stresses required to nucleate dislocations (**Fig. 3**), ruling out homogeneous dislocation nucleation at both the mesoscale and the macroscale. The discrete density of slip traces in **Fig. 2** and the higher yield strength in **Fig. 5**, particularly with Nb-doped (0.5 wt%), hint at a sluggish dislocation motion. Note, these stress levels are sufficient to activate dislocations in the $\{110\}_{45}$, contrary to the nanoindentation stress levels in **Fig. 3,** where dislocations in the $\{110\}_{90}$ (subscripts represent the inclination of the slip plane with the (001) surface [70]) directions are activated at shear stresses up to 10 GPa. The current authors have previously reported the resolved shear stress acting on the $\{110\}_{45}$ and $\{110\}_{90}$ dislocations [36], as 0.47 $p_o$ and 0.32 $p_o$ respectively [69]. Hence, it is much easier to activate dislocations in the $\{110\}_{45}$.

#### ***4.2.3. Dislocation multiplication***

Although dislocation multiplication events have been directly visualized in some crystalline solids [71], direct observation of dislocation multiplication is non-trivial in oxides. Multiplication events cannot be entirely decoupled from the motion of dislocations, as the dislocations need to move to multiply [72]. Hence, the interpretation in **Sec. 4.2.2** would complement the discussion here. The increase in slip trace density observed in **Fig. 2** after multiple cyclic Brinell indentation cycles is attributed to the multiplication of dislocations, similar to nominally undoped single-crystal $SrTiO_3$ [36]. However, a closer examination of the features in **Fig. 2** highlights the suppressed dislocation multiplication in the Nb-doped sample, particularly in the 0.5 wt% Nb-doped sample. Particularly, highlighted in the AFM images in **Figs. 2g & h**, where a one order of magnitude difference in the slip trace spacing is observed between the reference and Nb-doped (0.5 wt%) samples. An easier dislocation multiplication, such as a multiple cross-glide [72] mechanism, would be represented by the reference sample. This agrees with the significant decrease in the multiplication events in the 0.5 wt% Nb-doped sample. It is likely that, as strontium vacancies readily pin moving dislocations **(Fig. 7d)** and lower oxygen vacancy concentrations lead to less heterogeneous dislocation nucleation, the possibility for multiplication events is limited, leading to a discrete density of slip traces observed in **Figs. 2c & f**. Given the lack of an interatomic potential for Nb-doped $SrTiO_3$, this hypothesis is essential to test in future computational simulation works.

### 4.3. Correlation of the dislocation behavior across the length scales

Here, we summarize the impact of Nb-doping on the dislocation behavior of $SrTiO_3$ at different scales. First, we observed a scale-dependent competition influencing the deformation behavior at different length scales. Dislocation nucleation is facilitated by the oxygen vacancies in the reference sample compared to the Nb-doped samples, as illustrated in **Fig. 7e, f** by the higher population of oxygen vacancies within the probed volume for the reference sample. The findings were validated on the Fe-doped sample with equivalent doping concentration as the Nb-doped sample (**Fig. 8**), and supported by the previous experiments using electromigration coupled nanoindentation tests with MD simulation [43]. However, regardless of the vacancy types as schematically illustrated in **Figs. 7c, d**, (oxygen vacancy or strontium vacancy), the motion of dislocation is hindered by the vacancies, with strontium vacancies showing a more drastic effect on Nb-doped samples.

A significant shift in the dominating deformation mechanism is observed at larger length scales, as dislocation motion and multiplication dominate with increased stressed volumes and at lower stress levels [32] (mesoscale and macroscale). Contrary to the nominally undoped $SrTiO_3$, Nb-doped samples (strontium vacancy charge compensated) lead to a more severe suppression of dislocation motion, as schematically illustrated in **Fig. 7d**. Irrespective of the length scale, the dominating vacancy type dictates the behavior of strontium titanate under mechanical stress. These findings underscore the impact of defect chemistry on the plasticity of strontium titanate at room temperature.

### 4.4. Prediction and screening plasticity in other oxide ceramics

Point defect engineering can be used to tune the mechanical response of other oxides, provided the interaction of point defects with dislocations is well understood. For instance, the defect chemistry of $BaTiO_3$ is similar to that of $SrTiO_3$, though with a few approximations. In both perovskites, oxygen vacancies are more mobile than cation vacancies at low and intermediate temperatures [59]. Contrary to $SrTiO_3$, B-site vacancies in $BaTiO_3$ are more feasible under oxidizing conditions and donor doping [73]. Whether such differences translate into distinct mechanical responses remains an open question to be addressed in the near future.

Our findings indicate that nano-/microscale and mesoscale tests such as nanoindentation (**Figs. 3 & 4**) and cyclic Brinell indentation (**Fig. 2**) are practical tools for assessing dislocation nucleation, multiplication, and motion. The high cost and material constraints (e.g., availability of bulk crystals) associated with bulk compression of single crystals pose limitations and challenges. We have demonstrated that nano-/micromechanics combined with mesoscale testing is adequate to capture the dislocation mechanisms and predict the bulk compression behavior of single-crystal oxides. As this combination requires much less material, it is expected to help probe plasticity in other oxide ceramics and guide defect-engineering strategies in functional ceramics.

## 5. Conclusion

This study provides a systematic multiscale investigation of the impact of donor (Nb) doping on the room-temperature dislocation plasticity of $SrTiO_3$ via nanoindentation, Brinell cyclic indentation, and bulk uniaxial compression. Across the length scale, 0.5 wt% Nb doping consistently suppresses dislocation nucleation, multiplication, and motion, effectively regulating the dislocation-based plasticity of $SrTiO_3$. At the nano-/microscale, dislocation nucleation is facilitated by oxygen vacancies in the nominally undoped sample. In contrast, the dominant point defect species induced by Nb-doping (Sr vacancies) suppresses dislocation nucleation, as evidenced by the Fe-doped sample with an equivalent doping concentration. We attribute the higher lattice friction stress and lower creep strain rate to the sluggish diffusion of strontium vacancies in the Nb-doped samples. Although oxygen vacancies can also hinder the dislocation motion, the drag-like effect tends to scale with the diffusivity of individual vacancy types. These findings corroborate the observed discrete slip traces at the mesoscale and the higher yield strength at the macroscale for the 0.5 wt% Nb-doped sample. The results highlight the vital role of defect chemistry in tuning the mechanical properties of functional oxides at room temperature. By controlling the dopant/doping concentration and associated defect-defect interactions, it should be possible to engineer oxides with tailored mechanical response for coupled dislocation-point defect functional applications. Further studies could explore temperature-dependent interactions between dislocations and point defects, focusing on the thermal stability of these interactions for typical functional oxide applications.

**Acknowledgement**

C. Okafor and X. Fang acknowledge the financial support by the Deutsche Forschungsgemeinschaft (DFG, grant No. 510801687). X. Fang is also supported by the European Union (ERC Starting Grant, Project MECERDIS, grant No. 101076167). I. Huck acknowledges the support of the Deutsche Forschungsgemeinschaft (DFG, grant No. 501386284). Views and opinions expressed are, however, those of the authors only and do not necessarily reflect those of the European Union or European Research Council. Neither the European Union nor the granting authority can be held responsible for them. This work was partly carried out with the support of the Karlsruhe Nano Micro Facility (KNMFi, www.knmf.kit.edu), a Helmholtz Research infrastructure at Karlsruhe Institute of Technology (KIT, www.kit.edu). This work was partly carried out with the support of the Joint Laboratory Model and Data-driven Materials Characterization (JL MDMC), a cross-center platform of the Helmholtz Association. We thank Prof. Roger A. De Souza at RWTH Aachen for the insightful discussions.

**Author contributions:**

Conceptualization & Project design: XF; Methodology: CO, KT, SK, IH, AN, XF; Investigation: CO; Visualization: CO; Funding acquisition: XF, AN; Project administration: XF, AN; Supervision: XF, AN; Writing – original draft: CO; Writing – review and editing: CO, KT, SK, IH, SB, RL, YL, KD, AN, XF.

**Conflict of interest:**

The authors declare that they have no competing interests.

**Data and materials availability:**

All data are available in the main text or the Supplementary Materials